# *Hydrodynamic Simulations using GPGPU Architectures*

**Adrian COMAN, Elena APOSTOL, Cătălin LEORDEANU, Emil SLUȘANSCHI**
*University Politehnica of Bucharest, Faculty of Automatic Control and Computers*
*adrian.coman@cti.pub.ro, elena.apostol@cs.pub.ro, catalin.leordeanu@cs.pub.ro, emil.slusanschi@cs.pub.ro*

***Abstract:*** Simulating the flow of different fluids can be a highly computational intensive process, which requires large amounts of resources. Recently there has been a lot of research effort directed towards GPU processing, which can greatly increase the performance of different applications, such as Smoothed Particle Hydrodynamics (SPH), which is most commonly used for hydrodynamic simulations.

The introduction of GPGPU (General-Purpose Graphics Processing Unit) now allows a transition from large HPC clusters to workstations with graphic oriented devices. This new paradigm provides a cheaper alternative to researchers and practitioners to do numerical computations. Such applications developed for the GPU have used assembly and shader languages in the past (GLSL, HLSL). The recent rise in popularity has brought forth multiple frameworks that try to reduce the amount of development needed. Such frameworks include CUDA, NVIDIA's proprietary acceleration platform and the open standard OpenCL, developed by the Khronos Group and adopted by all major GPU manufacturers (NVIDIA, AMD, IBM, Intel). Both frameworks provide similar APIs to manage memory and launch execution of programs called kernels on an accelerator. In contrast to CUDA which can run only on NVIDIA's hardware, OpenCL is a heterogeneous solution that can run on any type of architecture, from CPUs, or GPUs to FPGAs.

Smoothed particle hydrodynamics (SPH) is a numerical method commonly used in Computational Fluid Dynamics (CFD). It is a method that can simulate particle flow and interaction with structures and highly deformable bodies. It replaces the fluid with a set of particles that carry properties such as mass, speed and position that move according to the governing dynamics. The dynamics of fluids are based on the Navier-Stokes equations. These describe the physical properties of continuous fields in the fluid. SPH approximates these equations using an integral interpolant that is then solved numerically.

This article addresses the current state of technologies available that can be used to speed up the algorithm and proposes a set of optimizations that can achieved by using different frameworks. We also draw conclusions regarding the equilibrium between performance and accuracy, using different numerical algorithms, frameworks and hardware optimizations.

## INRODUCTION

Smoothed particle hydrodynamics (SPH) *(Monaghan, 2005)* is a numerical method commonly used in computational fluid dynamics (CFD). Traditionally there have been two general approaches towards simulating the flow of fluids. The first one was the Eulerian method that uses geometric grids in which the fluid characteristics are determined for every cell. These were the foundation for codes build in the late 1960s with applications in many fields including oceanography, civil engineering, automotive and aerospace.

SPH uses the second approach, a Lagrangian mesh-less method that can simulate particle flow and interaction with structures and highly deformable bodies *(Cossins, 2010).* Originally used in computational astrophysics, SPH approximates values and derivatives in continuous domains by interpolating discrete sample points. It replaces the fluid with a set of particles that carry properties such as mass, speed and position. This has many advantages: it is an exact model for solving advection, it trivially solves interactions between different materials, can be easily modified to support molecular dynamics simulations and is computationally efficient in terms of storage and calculations as work is done only where matter is present *(Monaghan & Kos, 1999)*.

Particles in the system have an associated smoothing space in which a kernel function is applied. The kernel function smooths the properties of neighboring particles in the given space at each time iteration. This simulates the stronger effect nearby particles have compared to others that are at a greater distance. SPH in its weakly-compressible formulation is very computationally expensive because of the large domain and the high resolution at which it is solved. This has limited the time interval for which a simulation is solved or required the use of powerful hardware.

The SPH method is reaching maturity with constant improvements in accuracy and reliability. This paper presents the SPH numerical method and the governing equations that have led to it. It also provides a comparison of current free implementations and how they took advantage of the compute power of GPUs.

The article addresses the current state of technologies available that can be used to speed up the algorithm. The second chapter briefly looks at the underlying mathematical models and describes the SPH simulation problem. The third and fourth chapters provide an overview of different parallel frameworks used to accelerate these kinds of simulations and describe different available implementations. We look for potential optimizations that can achieve by using such frameworks. Existing software solutions are compared to see what features and performance they offer and how to leverage these new (or old) frameworks and hardware to gain speed and accuracy. Finally, we present the experiments we performed, draw conclusions and discuss further extensions and improvements to the algorithm.

**FLUID DYNAMICS**

The flow of Newtonian fluids with speeds much smaller than the speed of light is simulated using the Navier-Stokes equations. This non-linear set of differential equations describes how the field tension of the fluid varies with the flow velocity gradients and pressure. Implementations of the Navier-Stokes equations have followed two approaches: the Eulerian and Lagrangian method *(Kelager, 2006;* Al-*Fulaij et al., 2016)*.

**Navier-Stokes Equations**

The equations for conservation of momentum and continuity are

$$\frac{\partial \boldsymbol{u}}{\partial t} = -\frac{\boldsymbol{\nabla} P}{\rho} + v \boldsymbol{\nabla} \cdot \boldsymbol{u} + \text{g}$$

and

$$\frac{\partial \rho}{\partial t} + \rho \boldsymbol{\nabla} \cdot \boldsymbol{u} = \boldsymbol{0}$$

where $u$ represents velocity, $\rho$ is density, $v$ is viscosity, p is pressure and $g$ is gravity.

The first equation is Newton's second law for fluids as it describes how pressure, inertia and gravity affect the flow. The formulation is based on a mesh structure in which particle quantities vary with time as well as position. The right hand side represents the total forces as a product of mass-density and acceleration-density.

The continuity equation (or conservation of mass) can be simplified when density is constant ($\frac{\partial \rho}{\partial t} = 0$ and $\rho \boldsymbol{\nabla} \cdot \boldsymbol{u} = \boldsymbol{\nabla} u$). Thus, the equation for incompressible fluids becomes:

$$\boldsymbol{\nabla} \cdot \boldsymbol{u} = 0$$

**The SPH formulation**

SPH is a method based on approximating values in the continuum field with discrete particles that can exactly quantify physical properties such as mass, velocity, position etc. It uses the discretized Navier-Stokes equations which are locally integrate at each particle position. The quantities are weighted averages of neighboring particles, which are determined using a distance function characterized by a smoothing length h.

The partial differential equations of the conservation laws are interpolated to estimate values at discrete points in the field. The interpolation function is known as a *smoothing kernel* (W) and usually approximates a delta function. The integral describes any quantity function using points r' from the domain Ω.

$$A_I(r) = \int_\Omega A_I(r')W(r - r_b, h)dr'$$

The integral can be approximated by a summation over discrete particles in Ω as:

$$A_S(r) = \sum_b A_S(r_b)W(r - r_b, h)\Delta v_b$$

In practice the kernel function declines rapidly with distance and only neighboring particles are taken into account. $\Delta vb$ is the volume of particle b and can be rewritten as $m_b/\rho_b$ where m and ρ are the mass and density of particle b. The equation then becomes:

$$A_S(r) = \sum_b A_S(r_b)\frac{m_b}{\rho_b}W(r - r_b, h)$$

For example, if A is the density function ρ, one can deduce the interpolation formula that estimates density at point r as:

$$\rho(r) = \sum_b m_b W(r - r_b, h)$$

This shows how the masses of particles in an area are smoothed to approximate the density in a point. The first derivatives can be estimated easily if W is a differentiable function, to give the equation

$$\frac{\cdot \partial A_s}{\partial x} = \sum_b m_b \frac{A_b}{\rho b}\frac{\partial W}{\partial x}$$

In many SPH problems, the fluid domain is usually bound by different surfaces or contains solid object that interact with it. These boundaries can be modelled by using particles that have different properties and behaviour but are otherwise processed in the same way.

Free surface particles are found at the boundary between homogeneous fluids (eg. water and air). These particles must satisfy the dynamic boundary condition which states that particles will remain on the boundary surface. They also do not suffer from parallel shear stress and perpendicular normal stress (kinematic condition).

**PARALLELIZATION FRAMEWORKS**

Parallelization frameworks are used in High Performance Computing (HPC) to solve complex problems. These problems are computationally intensive and require a significant amount of resources, beyond the scope of what a normal consumer system could provide. HPC make use of clusters or supercomputers, entities that can handle and divide the problem and solve it on multiple compute instances.

Whether using the specialized hardware of a supercomputer or the interconnected nodes of a cluster, the software has to be specially designed to handle data transfer between nodes or processes and synchronize execution. Several frameworks have been developed to move the focus on the problem solving then system management when implementing this kind of applications.

OpenMP is an API for multi-platform shared-memory parallel programming. It is available in C, C++ and Fortran and provides an easier method of using programming execution and handling of threads. Directives are used to mark and specify how a section of the program is to be executed in parallel. The compiler then replaces the section with a fork-join construct that spawns threads and divides the tasks among them. Because it is implemented on top of threads, OpenMP is best used to optimized the CPU usage of multi-core machines.

GPGPU involves using the parallel computing power of GPUs for common programs. It focuses on two main parallel programming paradigms: data-parallel programming and task-parallel programming. A data-parallel model focuses on executing a single sequence of instructions concurrently on a data set which is usually structured as an array. The problem's data is mapped over compute units and concurrently updated by them. It also provides hierarchical data parallelism using work-groups, allowing group synchronization and fast communication between work-items through shared memory.

Task-parallelism appears in problems that benefit from the vector operations supported by the device or in a series of tasks that do not affect each other and could execute in an out-of-order queue.

CUDA is a proprietary API developed by NVIDIA for their GPU devices. It follows the same GPGPU paradigm as OpenCL, in which the computationally intensive part of an application is accelerated by parallelizing it across a large number of processing elements. These processing elements or CUDA cores have a similar hierarchical memory but the local memory can be split between shared memory or cache and their sizes adjusted. OpenACC provides directives similar to OpenMP that, when compiled, convert and parallelize code using CUDA.

OpenCL is a framework for programming heterogeneous platforms composed of CPUs, GPUs, FPGAs or other processors. It allows writing a single program that can run on any number of systems ranging from mobile phones to supercomputers. Programs are called kernels and are coded using a subset of the C language. The kernels are controlled from the host via an API and are compiled at runtime for the device they are executing on. OpenCL defines a four-level memory hierarchy for the compute device, although not all of them have to be present:

- global memory: main memory of the device, shared by all processing elements, high
- access latency;
- read-only memory: smaller, low latency, writable only by the host;
- local memory: shared by a group of processing elements;
- per-element private memory (registers).

**GENERAL OVERVIEW OF SPH IMPLEMENTATIONS**

A SPH simulation consists of solving the set of equations for mass and movement at discrete points in time. While some tasks are easily parallelized, a few problems arise due to the movement of the particles. The implementations must follow some strategies such as minimizing the data transferred between nodes and/or devices, optimize memory usage and access patterns, hide latency etc. The solver algorithm generally has three phases:

1) *Generating the neighbor list.* The fluid domain must be divided into cells to restrict the search space, because only the particles inside the kernel domain are relevant for the processed particle.

2) *Particle Interactions.* For every particle in the domain, the momentum and continuity equations are solved. These are computed using other particles that are in the same cell or in neighboring cells.

3) *Updating the system.* Using the values calculated in the previous step, the particle positions are adjusted for the next time step. This operation can differ depending on the choice of time integrator. Every few iterations, particle data is dumped to disk for further analysis and post-processing.

While some tasks are easily parallelized, a few problems arise due to the movement of the particles. The implementations must follow some strategies such as minimizing the data transferred between nodes and/or devices, optimize memory usage and access patterns, hide latency etc.

For the neighbor search stage, several algorithms have been proposed, each being relevant for different use cases and hardware. For example, in the case of cell-linked lists, the physical properties of particles are stored in arrays that can be sorted following the order of the cells. This insures good memory access patterns for CPU, because neighboring particles are grouped together. Some interactions are symmetrical, meaning that the forces between two particles are equal but opposing. The forces could be calculated for only one particle, reducing the computation time in half. These optimizations are not applicable to GPUs as they run into memory race conditions.

Generating an expanded neighbor list (Verlet List) can have the advantage that future neighbors are also included. This acts as a cache and means that the search can be done once every few iterations. Most simulations will be in 3D, so solving the particle interactions for each direction can be done in parallel using SIMD instructions if the hardware supports it. This allows basic operations to be done simultaneously for four numbers. Because the kernel takes into consideration only particles in a sphere radius of h, the usual cell size (2h x 2h x 2h) can be reduced to improve the ratio between the cell and kernel domain volumes.

Common optimizations specific to GPUs *(Dalrymple et al., 2011)* include reducing global memory access and maximizing the number of active threads. These have to be balanced against each other. Faster memory access can be achieved by bringing that closer to the processing elements, in registry or local memory. The number of threads depends on how many registries are required per thread, since they are a limiting resource of the work-group.

**Smoothing Kernel Radius**

Choosing the support radius h is critical for an accurate simulation. It affects the number of particles that are used when integrating the property values of a point. Although a larger radius would seem to increase the precision, this is generally not a good idea because of the way the kernel works. Closer particles are weighted less than with other kernel lengths while further ones would have a great enough impact to influence the accuracy. When h→0, too few particles are taken into consideration and the results are again imprecise.

The radius is determined as a function of the number of particles that have to fill the kernel space, considering a constant density. The chosen number of particles (k) depends on the nature and properties of the fluid. If the support radius is considered as a sphere, then its length can be calculated as:

$$h = \sqrt[3]{\frac{3Vk}{4\pi n}}$$

**Finding the nearest neighbors**

All particles must evaluate several equations, each one directly dependent on the neighboring particles. A naive approach would be to search all particles in the system to see if they are in range. This is a very expensive method due to the time complexity being O($n^2$).

Faster search algorithms divide the space into voxels or cells which limit the number of particles that are of interest. These cells must be big enough to contain the kernel space. The time complexity then decreases to O(n · m) where m is the average number of particles found in a cell and its adjacent neighbors.

**ANALYSIS OF SPH SOFTWARE**

This chapter briefly presents current SPH software, the features and performance they offer as well as the technologies used in their implementation architecture. Some post-processing tools for visualization and analysis are also mentioned.

**SPH Solvers**

AQUAgpusph *(Cercos-Pita, 2015)* is an open source project that utilizes OpenCL *(Tompson & Schlachter, 2012)* to solve the Navier-Stokes equations using the weakly compressible SPH approach. It assumes that the fluid is barometric and will have only small variations of density throughout its domain. It is based on a similar model to the client-server architecture. The client has the job of initializing the environment and parsing the initial conditions. After that, requests are generated to the server to compute the physical properties and advance the system in time. The output is received and written to disk for post-processing. This procedure is repeated until the entire time interval is solved and the client shuts down the server. The server is related to the OpenCL concept of platforms and offers the possibility of horizontal scaling. This feature of adding more servers is not currently supported. OpenCL was chosen due to its ability to run on multiple platforms, as opposed to CUDA. It can cover all usage scenarios: serial and parallel CPU as well as a GPU version. It was used to implement the most popular boundary conditions as well as allow easy customization of the code due to the runtime compilation of the kernel. Boundary conditions are modelled as fixed or ghost particles, with the ability to also add elastic bounce for more complex geometries.

AQUAgpusph *(Cercos-Pita, 2015)* is Python extensible, to allow more complicated scenarios with moving solids to be simulated. The scripts are executed on the server alongside the OpenCL kernel.

SPHysics was an SPH solver written in FORTRAN. It was release in 2007 and since then it has evolved into versions for different platforms. One of these versions is DualSPHysics *(Alonso, 2014; Crespo et al., 2011; Crespo et al., 2015)*, a CUDA accelerated SPH solver that was validated for many problems of wave-breaking, dam-break and coastal structure interactions.

Initial conditions for the simulation are provided using XML files. These hold all the configuration and execution parameters: gravity, smoothing size, particle numbers, domain geometry etc.

GPUSPH was inspired by SPHysics and was first to implement GPU support for the SPH algorithm. Like DualSPHysics, it is built on top of CUDA and offers pretty much the same features. It doesn't offer the same level of ease of use as changing the parameters of a problem requires recompiling the software.

Another framework for SPH simulations is PySPH *(Puri et al., 2014)*. It pushes toward becoming a flexible, user friendly tool for prototyping new approaches to SPH or extending existing ones. The solver supports variable smoothing lengths, dynamic and repulsive boundary conditions for free surface flows and multi-stage time integrators.

PySPH is a hybrid framework written in Python. It uses Cython to improve performance of compute intensive sections by generating C extensions. These can use MPI as well as OpenCL to execute tasks simultaneously. There are 4 modules that together form the software package:

- base defines all data structures for particles, oct-trees, neighbor lists (Dominguez et al., 2011)
- sph provides an interface to all physical interactions that can be computed
- solver is the main package that models the SPH solution using physics from sph
- parallel is responsible for scattering data to the processing entities, provide shared resources and balance the load. It uses a custom python wrapper over Zoltan.

The Zoltan library is a toolkit that provides the dynamic load-balancing and data partitioning algorithms. Another module called Mayavi can handle visualizations as well as control the simulation during execution. This allows for tweaking values mid run. Output can also be done in VTK format or for python scientific tools such as matplot or numpy.

**Visualization toolkits**

VTK is an open-source, freely available software system for 3D computer graphics, image processing, and visualization. The visualization algorithms include scalar, vector, tensor, texture, and volumetric methods, as well as advanced modelling techniques such as implicit modelling, polygon reduction, mesh smoothing, cutting, and contouring. Various implementations of VTK exist, the most notable of which is ParaView. It allows interactive 3D exploration of the data, filtering by units of interest and a layered separation of data types.

**EXPERIMENTAL RESULTS**

The most iconic test scenarios for testing SPH code is the dam break one. The test case consists of a 3D dam break flow hitting a structure. No moving boundary particles are involved in this simulation and no movement is defined. The physical time simulated is 1.5 seconds, *(Fig.1, 2)*.

This scenario was used to test AQUAgpusph, DualSPHysics and GPUSPH with different resolutions on a GTX 960 with 1024 CUDA cores and 2GB GDDR5. The results are shown in the *Figure 3*.

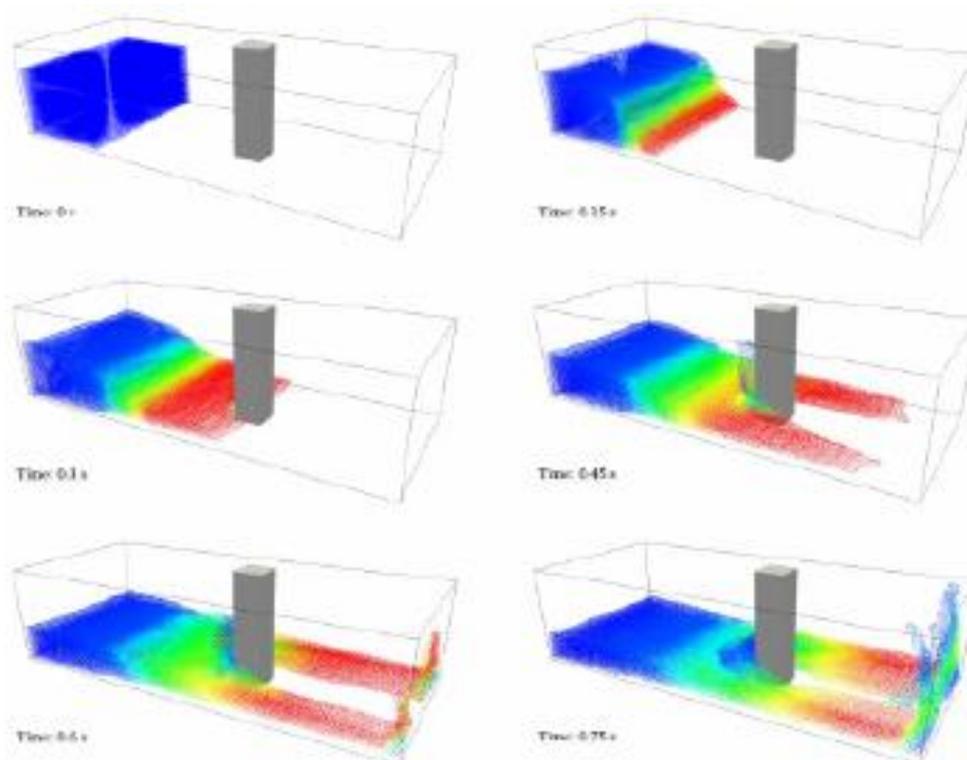

*Figure 1. 3D dam break at different points in time. Red particles have higher velocities than blue ones.*

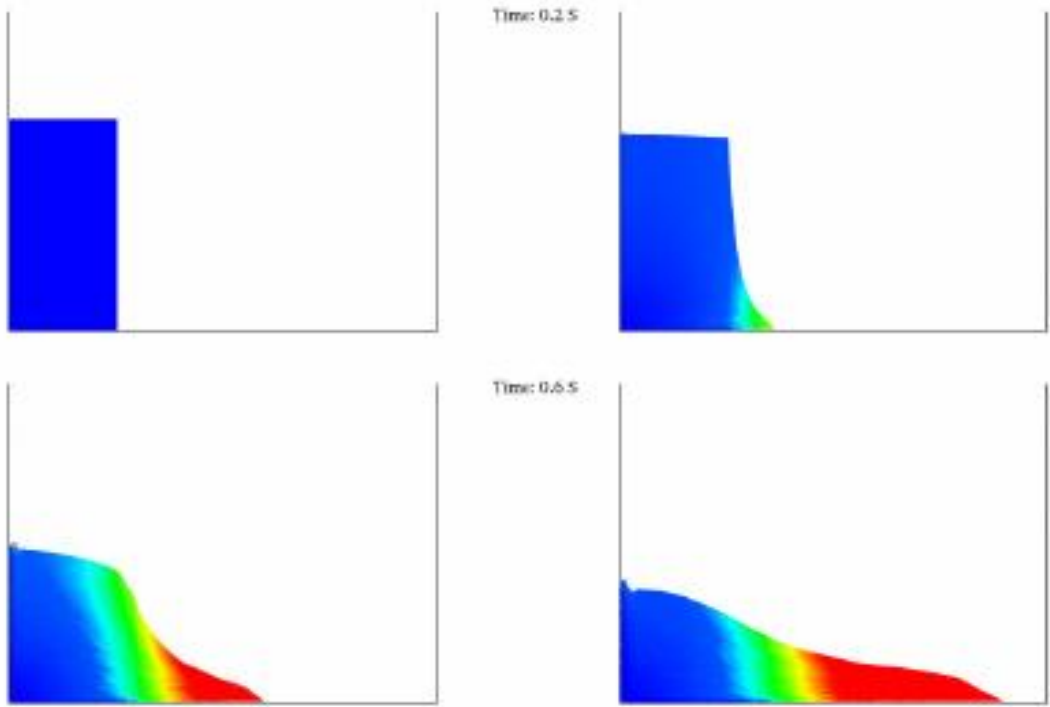

*Figure 2. 2D slice of a dam break*

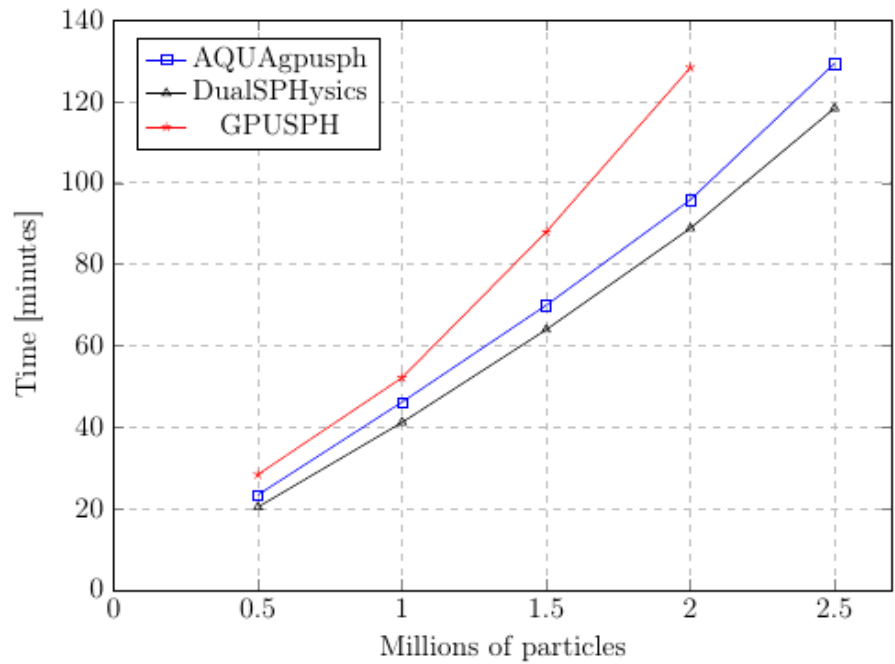

*Figure 3. Application run times for different resolutions*

**CONCLUSIONS**

This paper discussed the programming paradigms that can be used to speed up the SPH algorithm. Real time interactive simulations are a desirable feature for certain industries and researchers.

Previously, the algorithm was heavily criticized due to being computationally expensive, requiring days to run on serial or parallel CPUs or entailing significant investments in supercomputers or clusters. The new GPU paradigms have evolved from the times when code would have to be written in assembly or as graphic shaders.

Current frameworks such as OpenCL and CUDA provide great performance improvements while lowering the price, while offering a more accessible learning curve. The solutions presented here are almost evenly matched in performance, but each offer slightly different features and platform support. They are also still in active development, with upcoming features and updates. AQUAgpusph provides a nice balance between cross-platform support and performance, achieving almost similar performance to DualSPHysics which can only run on NVIDIA accelerators. PySPH offers a sandbox for prototyping new approaches when solving the governing equations, while partitioning the fluid domain more efficiently.

All solutions discussed are open-source and available for further development. Most of them are missing multi-GPU support or any other means of scaling across multiple devices.